\documentclass[hyper,11pt,letterpaper]{JHEP3}
\usepackage[dvips]{epsfig}
\usepackage{amsmath,amssymb,epsf,amsfonts}
\usepackage{graphicx}
\usepackage{dcolumn}
\usepackage{bm}

\usepackage[english]{babel}
\usepackage{amssymb}
\usepackage{amsfonts}
\usepackage{amsmath}
\usepackage{graphicx}
\usepackage{epsfig}
\usepackage{cite}
\newcommand{\be}{\begin{equation}} \newcommand{\ee}{\end{equation}}
\newcommand{\bea}{\begin{eqnarray}} \newcommand{\eea}{\end{eqnarray}}
\newcommand{\beann}{\begin{eqnarray*}}  \newcommand{\eeann}{\end{eqnarray*}}
\newcommand{\bfig}{\begin{figure}} \newcommand{\efig}{\end{figure}}
\newcommand{\ba}{\begin{array}} \newcommand{\ea}{\end{array}}
\newcommand{\bcen}{\begin{center}} \newcommand{\ecen}{\end{center}}
\newcommand{\btab}{\begin{tabular}} \newcommand{\etab}{\end{tabular}}

\newtheorem{Proposition}{Proposition}[section]

\newtheorem{Theorem}{Theorem}[section]
\newtheorem{Lemma}{Lemma}[section]
\newtheorem{Corrolary}{Corrolary}[section]

\newcommand{\bp}{\begin{Proposition}}   \newcommand{\ep}{\end{Proposition}}
\newcommand{\bt}{\begin{Theorem}}   \newcommand{\et}{\end{Theorem}}
\newcommand{\bl}{\begin{Lemma}}     \newcommand{\el}{\end{Lemma}}
\newcommand{\bc}{\begin{Corrolary}} \newcommand{\ec}{\end{Corrolary}}
\title{ ABJM Mirrors and a Duality of Dualities}

\author{
Kristan Jensen\footnotemark[1]\,
and
Andreas Karch\footnotemark[2]
\\
Department of Physics, University of Washington, Seattle, WA
98195-1560}

\footnotetext[1]{E-mail: \email{kristanj@u.washington.edu}}
\footnotetext[2]{E-mail: \email{karch@phys.washington.edu}}

\abstract{
We clarify how mirror symmetry acts on 3d theories with ${\cal N}=2,3$ or 4 supersymmetries and non-abelian Chern-Simons terms and then construct many new examples. We identify a new duality, geometric duality, that allows us to generate large families of gauge theories, with and without Chern-Simons term, that all flow to the same conformal field theory in the infrared. In particular, we find an interesting duality of dualities: a pair of theories related via mirror symmetry can be mapped, via geometric duality, into a pair of gauge theories related by Seiberg duality.  This network of dualities can be understood as the
simple result that all of these theories are
different realizations of one and the same system in M-theory.
}

\begin{document}

\section{Introduction}

Supersymmetric gauge theories in three dimensions are known to exhibit a very
interesting duality: mirror symmetry \cite{Intriligator:1996ex}. Two different gauge theories flow to the same interacting conformal field theory in the infrared. Both theories have a moduli space of vacua. Mirror symmetry maps the Coulomb branch of one of the two theories to the Higgs branch of the other. Mirror symmetry is a property for both abelian and non-abelian gauge theories. In the ${\cal N}=4$ supersymmetric
case, abelian mirror symmetry can be understood in terms of a single path integral identity \cite{Kapustin:1999ha}. If true, all examples of abelian mirror symmetry can be derived from this identity. As pointed out e.g. in \cite{Sachdev:2008wv}, supersymmetric gauge theories in three dimensions are useful toy models for studies of condensed matter systems. They exhibit tractable examples of quantum criticality; mirror symmetry is the particle/vortex duality also known from the condensed matter literature \cite{Aharony:1997bx}. Another feature of mirror symmetry is the enhancement of global symmetry at the infrared fixed point \cite{Intriligator:1996ex}. Mirror symmetric theories typically have different manifest global symmetry groups, while the conformal field theory they flow to exhibits both. This enhanced symmetry typically does not have a description in terms of the fundamental fields of either theory and only occurs at the fixed point. We will make great use of this enhanced symmetry in this work.

Interest in 3d supersymmetric gauge theories has recently seen a renaissance, mostly due to the work by ABJM \cite{Aharony:2008ug}. In there, a particular example of an ${\cal N}=6$ supersymmetric gauge theory with Chern-Simons (CS) terms is found to have an interesting AdS/CFT dual. A natural follow-up is to ask what theories are mirror symmetric to ABJM?  More generally, we would like to answer the question: how does mirror symmetry act on theories with CS terms? Already in the early days of mirror symmetry it was realized that mirror symmetry often links theories with and without CS terms to each other \cite{Tong:2000ky}. However, a coherent picture for how mirror symmetry acts on a general theory with CS terms is unknown.
In the abelian case, mirror symmetry typically takes a theory at level $k_{CS}$ to a theory at level $1/k_{CS}$ \cite{Kapustin:1999ha}. In the non-abelian case the level has to be integer, so clearly mirror symmetry has to act differently.

Our starting point for these investigations is the realization of these gauge theories via brane setups in type IIB string theory.  To uncover the full network of dualities, we must also study D2 branes in type IIA string theory. All of these various brane realizations give very different gauge theories which in the end can be shown to belong to one and the same universality class. That is, they all flow to the same interacting CFT in the IR. In terms of string theory, all of the different brane setups lift to one and the same system in M-theory.

The outline of this paper is as follows. In Section \ref{n4mirror} we review the construction of 3d ${\cal N}=4$ supersymmetric gauge theories via brane configurations and how mirror symmetry acts on them with a special emphasis on the role of global symmetries. In Section \ref{mSymIIB} we review the brane construction for theories with CS terms. We show how mirror symmetry acts on these configurations. We show that for non-abelian theories with CS level bigger than one, we can still find a mirror. The mirror is defined via the gauging of an enhanced global symmetry of a known conformal field theory, albeit one without a Lagrangian description. We also present many new examples of mirror symmetry with CS terms by considering general $SL(2,\mathbb{Z})$ transformations.  In Section \ref{oneM} we note that many of the brane setups we consider lift to the same system in M-theory, namely M2 branes at the singularities of certain toric Calabi-Yau four-folds.  These four-folds are really product spaces of the form $CY_3\times \mathbb{R}^2$.  There are then at least two non-trivial reductions to a IIA description obtained by
reducing on either the trivial circle inside
$\mathbb{R}^2$ (after compactifying to $\mathbb{R}^1 \times S^1$)
 or reducing on a nontrivial circle inside of the $CY_3$.
Reducing on the trivial circle yields a setup with
D2 branes probing a toric $CY_3$ with no CS terms in the field theory, while reducing on the nontrivial circle leads to a theory with CS terms \cite{Aganagic:2009zk}.  We therefore find a new duality, geometric duality, that these different descriptions are equivalent.  In Section \ref{dualDualities} we then consider toric duality in the IIA description of these brane systems.  The toric duality of 3d gauge theories is well-known to be equivalent to Seiberg duality of their 4d parents, but we can take this a step further and, via geometric duality, map toric duality to mirror symmetry in the IIB brane setup.  All in all, this network of dualities simply realizes the fact that
these related theories are described by the same system in M-theory.

\section{Mirror symmetry from IIB brane setups}
\label{n4mirror}

Ever since the pioneering work of Hanany and Witten \cite{Hanany:1996ie},
the mirror symmetry of many 3d gauge theories has been understood as arising from
S-duality acting on an $\mathcal{N}=4$ brane configuration in IIB string theory involving
NS5 branes (along the 012345 directions) D3 branes (along the 0126 directions)
and D5 branes (along the 012789 directions). These brane constructions engineer
various quiver gauge theories with fundamental matter. S-duality leaves
the D3 branes untouched, but exchanges NS5 and D5
branes\footnote{When we perform an S-duality we also
automatically relabel the 345 directions as 789 and vice versa. 
As pointed out in the original \cite{Hanany:1996ie}, this action 
exchanges the two $SU(2)$ factors of the R-symmetry, as expected from mirror symmetry.}.

Of particular interest are the so called elliptical models, in
which the 6-direction (and thus the quiver) is circular. Let us for now focus on the
case where the number $N$ of D3 branes is the same between any
pair of five-branes. In this case the gauge group described by the brane
setup is simply $U(N)^k$ with $F$ flavors of fundamental
hypermultiplets, where $k$ is the number
of NS5 branes and $F$ the total number of D5 branes. In
addition to the fundamental hypermultiplets there
are also bifundamental hypermultiplets connecting the $i$-th
and the $(i+1)$-th gauge group factor. In the case where there is only one NS5 brane ($k=1$), this bifundamental matter is really an adjoint. Similarly, in the case with no NS5 branes ($k=0$), we are still describing a single $U(N)$ gauge
group with an adjoint hypermultiplet and $F$ flavors. In fact,
the gauge theories for $k=0$ and $k=1$ are identical (their mirrors however
are not).  Both of these setups at low energies describe $\mathcal{N}=8$ $U(N)$ super Yang-Mills coupled to $F$ fundamental hypers.

The $F_i$ fundamental flavors ($i=1,2,\ldots,k$, $\sum_i F_i =F$)
associated with the $F_i$
D5 branes lying between the $i$-th and $(i+1)$-th NS5 brane are
associated with the $i$-the gauge group factor. The global symmetry
rotating the flavors is $\prod_i U(F_i)$. The inverse gauge coupling
of the $i$-th factor is proportional to the length of the $i$-th
interval. In the deep IR all gauge couplings effectively become infinite,
the 6-circle shrinks to zero size and the theory flows to
an interacting CFT in the IR. An example of such a setup with $k=0$
and $F=8$ is displayed in panel A) of Figure (\ref{hw}). The gauge
group is ${\cal N}=4$ supersymmetric $U(N)$ with eight fundamental
hypers and one adjoint, i.e. $\mathcal{N}=8$ super Yang-Mills coupled to eight flavors.

\begin{FIGURE}
    {
\centering
    \includegraphics[scale=.53]{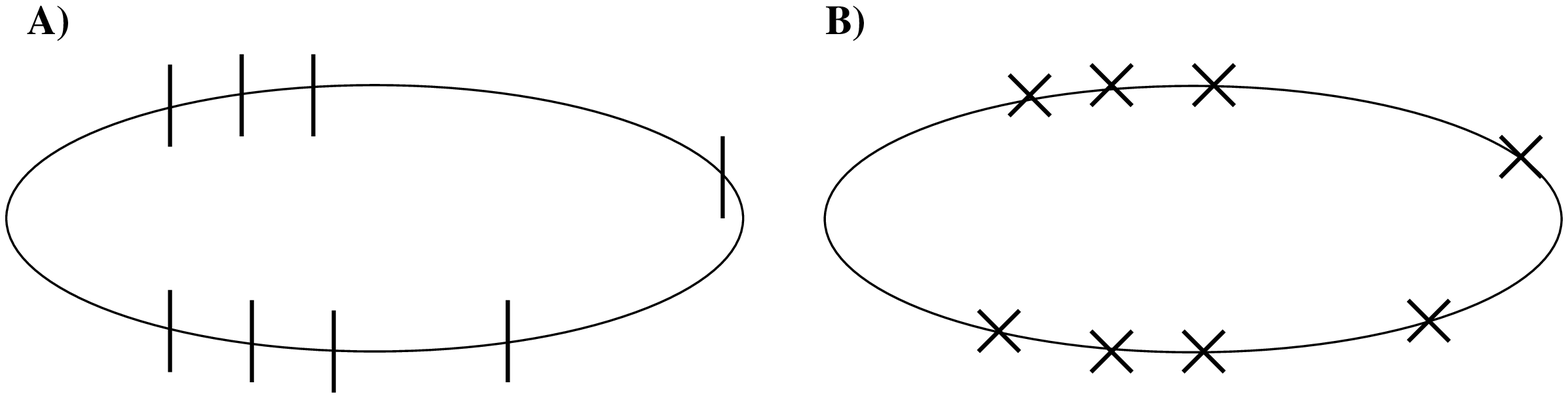}
    \caption{
    \label{hw}
Mirror symmetry for an elliptic
model with $k=0$ and $F=8$. Crosses indicate NS5 branes, while vertical
lines are D5 branes.
Panel A) describes the electric theory and panel B) its mirror.
     }
}
 \end{FIGURE}

The S-dual brane configuration is of exactly the same type, with NS5
and D5 branes exchanged. The mirror gauge group is $U(N)^F$ with
$k$ flavors and a global flavor symmetry $\prod_a U(k_a)$ (where,
similar to above, $a=1,2,\ldots, F$ and $\sum_a k_a=k$). The
brane configuration for $k=8$, $F=0$ is displayed in panel B)
of Figure (\ref{hw}). The gauge group is $U(N)^8$ with bifundamental
matter. There are no fundamental flavors in this case. Deep in the
IR, this field theory is believed to flow to the identical fixed
point as the original theory. For this to be possible, we need
the global symmetry on both sides to be enhanced
to $\prod_i U(N_i) \times \prod_a U(k_a)$. Indeed the theory
of $k_a$ coinciding NS5 branes is an $\mathcal{N}=(1,1)$ supersymmetric 6d
$U(k_a)$
gauge theory. From the point of view of the 3d field theory, this appears as
a global symmetry. This enhanced global symmetry at the IR fixed point
is one of the hallmarks of 3d mirror symmetry. A $U(1)^{k}$ subgroup of this enhanced global
$\prod_a U(k_a)$ symmetry can be explicitly seen in the original electric theory. For each photon from the abelian
$U(1)$ in each $U(N)$ gauge group factor, one can define a current $j_{\mu}=\epsilon_{\mu \nu \rho} F^{\nu \rho}$ which is trivially conserved as its divergence vanishes identically. The enhancement of this symmetry to
$\prod_a U(k_a)$ only occurs at the IR fixed point. For a more detailed discussion see the original mirror
symmetry paper \cite{Intriligator:1996ex}, or \cite{Aharony:1997bx,Kapustin:1999ha,Borokhov:2002cg}.
Away from the fixed point the Coulomb branch of one theory maps to the
Higgs branch of the other. Theories with ``mostly matter" ($F \gg k$) get
mapped into theories of ``mostly glue;" the former have a high-dimensional
Higgs branch and the latter a high-dimensional Coulomb branch.

The elliptical mirrors we just described also have a nice realization in
terms of type IIA string theory. In the T-dual language\cite{Ooguri:1995wj,Gregory:1997te,Tong:2002rq}, we
are describing $N$ D2 branes probing a $\mathbb{Z}_k$ singularity
(acting on the 6789 space) in the presence of $F$ D6 branes (along 0126789).
The ordering of the $F$ flavor branes into groups $F_i$ is achieved
by discrete Wilson lines: due to the orbifold action, the
Wilson line $e^{i\int A}$ of a single D6 worldvolume gauge field along
a path around the orbifold fixed point at $x_6=x_7=x_8=x_9=0$ can
take on any value $\omega^i$, where $\omega = e^{2 \pi i/k}$. A path
that winds $k$ times around the fixed point can be contracted to a point
and hence the $k$-th power of the Wilson line should vanish. We then identify the number of D6 branes with Wilson line $\omega^i$ as $F_i$. The non-abelian Wilson line of the $U(F)$ gauge field on the stack of $F$ D6 branes then block-diagonalizes into blocks with size $F_i$, realizing the flavor symmetry breaking $U(F)\rightarrow \prod_i U(F_i)$.

The advantage of the IIA picture is that it makes it manifest why both theories
flow to the same fixed point in the IR. The IR limit of
D2 branes is described by M2 branes in 11d M-theory. However, in
11d both D6 branes and the $\mathbb{Z}_k$ singularity lift to equivalent, purely
geometric singularities \cite{Porrati:1996xi}. The two different IIA descriptions
are just reductions along different $U(1)$ fibrations
of one and the same manifold. This is best understood
in the case of a ``purely geometric" (no D6s) theory and its ``branes
only" (no $\mathbb{Z}_k$ singularity) mirror. In the former case, we have $N$ D2s on a $\mathbb{Z}_k$ singularity,
the mirror has the D2s together with $k$ D6 branes. Either way,
in 11d these lift to M2 branes probing a $\mathbb{Z}_k$ singularity.
The corresponding field theory is simply an orbifold of the well
known M2 brane CFT. At
large $N$ this CFT has a field theory description in terms of
$AdS_4 \times S^7/\mathbb{Z}_k$.

The general case maps to $N$
M2 branes probing a $\mathbb{Z}_k \times \mathbb{Z}_F$ singularity \cite{Witten:2009xu}, with the first
factor acting on 6789 and the second on 345 and 10.
In the M-theory language one can again turn on discrete
fluxes in order to account for
the integers $F_i$ and $k_a$. At
a $Z_F$ fixed point, one studies 7d $U(F)$ SYM on $\mathbb{R}^{2,1} \times \mathbb{C}^2/\mathbb{Z}_k$;
at the $Z_k$ fixed point, 7d $U(k)$ SYM on $\mathbb{R}^{2,1} \times \mathbb{C}^2/\mathbb{Z}_F$.
For the first configuration, one can turn on $\binom{F+k-1}{F}$ different non-trivial Wilson lines at infinity (since $H_1(S^3/\mathbb{Z}_k)=\mathbb{Z}_k$), exactly counting
the number of different configurations of D5 and NS5 branes we started
with in the IIB setup. To see that we get the same counting, recall that there are a total of $(F+k)!$ orderings of branes on
the circle, but only $(F+k-1)!$ of them are inequivalent after
taking into account cyclic permutations. We can
always pick one of the NS5 branes to be the start of the first
interval, and then only have to distribute the remaining
$k-1$ NS5 branes and $F$ D5 branes. As all D5s and
all NS5s are indistinguishable among themselves, we have to
divide by $F!$ and $(k-1)!$ to only count once all configurations
that can be obtained from each other by permuting only D5
branes or only NS5 branes. There are hence $\binom{F+k-1}{F}$ ways to distribute the $F$ flavor D5 branes among
the $k$ intervals.

From the M-theory point of view, the two
singularities appear on an equivalent footing, and so there appears to be another set of discrete fluxes
associated with the
7d $U(k)$ SYM on $\mathbb{R}^{2,1} \times \mathbb{C}^2/\mathbb{Z}_F$. 
In \cite{Witten:2009xu} Witten shows that these two sets of fluxes on the two different singularities map to the ``linking numbers''
of the D5 branes and NS5 branes respectively, that is they encode the relative orderings of D5 and NS5 branes. However, from the type IIB setup we started with it is clear that this second set of discrete fluxes is not
independent \cite{Witten:2009xu}. Once
the D5 linking numbers are known (that is in our language the $F_i$), the NS5 brane linking numbers (or equivalently the $k_a$) are completely determined. So the two sets of discrete fluxes are not independent.

Let us close this section with a discussion of the case where
the number of 3-branes $N$ is not the
same on all intervals.
As our analysis easily extends to that
case, let us briefly describe what changes. Instead
of a $U(N)^k$ gauge group, we can have $k$ different gauge
group factors $\prod_i U(N + m_i)$, where $N+m_i$ is
the number of D3 branes stretched on the $i$-th interval.
We take $N$ to be the number of D3 branes on the interval
with the smallest number of D3s, so that $m_i \geq 0$ for all
$i$.
Under T-duality to the IIA picture the extra
branes map to fractional branes of the $\mathbb{Z}_k$ orbifold
\cite{Karch:1998yv}, that is D4 branes wrapping
the vanishing 2-cycles \cite{Douglas:1996sw}. In the
presence of the $F$ D6 branes these $m=\sum_i m_i$ D4
branes can simply dissolve into $m$ units of
worldvolume flux. In the M-theory lift, the fractional branes are again
represented by a discrete flux, but this time it is a discrete 4-form flux corresponding to the non-trivial torsion 3-cycles of $S^7/\mathbb{Z}_k\times \mathbb{Z}_F$ \cite{Aharony:2008gk}. These fluxes are different than the discrete fluxes we earlier associated with the breaking of global symmetry. The choices we found earlier were Wilson lines of the vector multiplet propagating at an orbifold fixed plane, as opposed to 4-form flux in the full transverse space. 

\section{Mirror symmetry for ${\cal N}=2$ and ${\cal N}=3$ CS-matter theories}
\label{mSymIIB}
\subsection{Brane construction}
\label{branesIIB}
In \cite{Kitao:1998mf,Bergman:1999na} it was shown that if one replaces a single NS5 brane with a ($k_{CS}$,1) brane
(that is a bound sate of $k_{CS}$ D5 branes with one NS5 brane), the gauge theory living on the interval to the left picks up a Chern-Simons term of level $k_{CS}$; similarly, the gauge theory in the interval on the right picks up a Chern-Simons term of level $-k_{CS}$. For the $(k_{CS},1)$ brane to preserve any remaining supersymmetries, one needs to rotate the brane in the 345789 space. One option is to rotate the brane in the 37 plane by an angle that is fixed in terms of $k_{CS}$. Out of the original $SO(3) \times SO(3) = SO(4)$ R-symmetry rotating the 345 and 789 space respectively, only an $SO(2) \times SO(2)$ is preserved, and thus supersymmetry is broken from ${\cal N}=4$ to ${\cal N}=2$. In our discussion below we will mostly focus on this generic ${\cal N}=2$ scenario. If one chooses to rotate by the same angle in the 37, 48 and 59 planes respectively, one can preserve
the diagonal $SO(3)$ of the original $SO(3) \times SO(3)$ and hence ${\cal N}=3$ supersymmetry.  The special case with $k=2$ (one NS5 and one $(k_{CS},1)$ brane) is the celebrated ABJM theory \cite{Aharony:2008ug}. In that case, the supersymmetry is further enhanced to ${\cal N}=6$ in the IR (or even ${\cal N}=8$ for $k_{CS}=1$), but this is not the case for any $k > 2$.

This construction opens the door for analyzing mirror symmetry in this context. S-duality takes a $(k_{CS},1)$ brane
into a $(1,k_{CS})$ brane, so unless $k_{CS}=1$ or $N=1$ one needs to understand the rules governing the brane setup in that case. The naive generalization of the results of \cite{Kitao:1998mf,Bergman:1999na}, a theory with a Chern-Simons terms of level $1/k_{CS}$, is clearly incorrect, as the level of a non-abelian Chern-Simons term must be integer in order to be gauge invariant. The arguments presented in \cite{Kitao:1998mf} would give level $p/q$ for any $(p,q)$ brane, which also clearly fails when $q\geq 2$. The derivation of \cite{Bergman:1999na} makes it clear why the case of a $(p,q)$ brane with $q\neq1$ (that is more than one NS5 brane in the boundstate of $p$ D5s and $q$ NS5s) is different and also gives an in-principle construction of the corresponding field theory. Let us briefly review this argument.

The crucial new ingredient in ${\cal N}=2$ theories is a D5' brane, that is a D5 brane along 012457. This D5' brane together with
all the other branes introduced above still
preserves ${\cal N}=2$ supersymmetry.
We will also need its S-dual, the NS5' brane along
012389. Adding $F'$ D5' branes introduces  fundamental flavors as with
the D5 branes.  In the language of $\mathcal{N}=2$ multiplets, each D5' brane adds one fundamental chiral field $Q$ and one anti-fundamental chiral field $\tilde{Q}$. Unlike the case with D5 branes, the $X Q \tilde{Q}$ superpotential term demanded by ${\cal N}=4$ supersymmetry (coupling the adjoint chiral $X$ from the ${\cal N}=4$ vector multiplet to $Q$ and $\tilde{Q}$) is absent. This has a major consequence: the global flavor symmetry is not just the $\prod_i U(F_i)$ we had for $F_i$ D5 branes on the $i$-th interval, but instead we have a global $\prod_{i} U(F'_i) \times U(F'_i)$ (where again $F'_i$ denotes the number of D5' branes on the $i$-th interval and hence $\sum_i F'_i = F'$). Without the superpotential term, one can rotate the $Q$ and $\tilde{Q}$ flavors independently.

Under S-duality, D5' branes turn into NS5' branes. The rules governing NS5' branes are similar to the ones of NS5 branes. In particular, the gauge group will still be $U(N)^{k+k'}$, where $k$ still denotes the total number of NS5 branes and $k'$ the total number of NS5' branes. Whenever an interval is bounded by two NS5 branes or two NS5' branes, the corresponding gauge group factor is an ${\cal N}=4$ $U(N)$ theory, that is in the ${\cal N}=2$ language a $U(N)$ gauge theory with an adjoint hypermultiplet $X$.
However, when one side is an NS5 and the other an
NS5' $X$ gets a mass and can be integrated out\footnote{A quartic superpotential is also generated. The rules for finding the
correct superpotential for the bifundamentals for a generic series of NS5 and NS5' branes
have been worked out in \cite{Uranga:1998vf}. For the flavors from D5 or D5' branes associated
with a given NS-NS' interval one also generates a quartic superpotential, the details of which dependd on the ordering of D5 and D5' branes \cite{Aharony:1997ju}.}.
One is left with an ${\cal N}=2$ $U(N)$ gauge group factor. The flavors associated with D5 branes have a cubic
superpotential with $X$ when suspended between NS5 branes,
but not when between NS5' branes. The opposite is true for D5' brane flavors.
A straightforward
generalization of this construction involves branes that are at different angles in the 48 and 59 plane than the ones we described above. Rotating a D5 brane from a D5 into a D5' in these planes continuously turns off the cubic superpotential, while rotating an NS5 to NS5' increases the mass of $X$ from 0 to infinity.

 Note that, while replacing D5 branes with D5' branes in the electric theory gives an enhancement of the global symmetry from $\prod_i U(F_i)$ to $\prod_i U(F_i) \times U(F_i)$, the mirror gauge theory still
only has a $U(1)^F$ manifest global symmetry from the $F$ dual photons of the mirror symmetry. At the fixed point, of course, the mirror has to have the full enhanced global symmetry. This means that for the ${\cal N}=2$ theory of $k'_a$ coinciding NS5' branes in the presence of D5 branes (the mirror of an electric theory with NS5 branes and D5' branes only) the global symmetry is enhanced not just from $U(1)^{k'_a}$ to $U(k'_a)$ but instead to $U(k'_a) \times U(k'_a)$. From the brane point of view, this enhanced symmetry is manifest in both the electric and the mirror theory. Since D5' branes and NS5 branes share five worldvolume directions (01245), they can form a ``$(p,q)$-web" \cite{Aharony:1997ju,Aharony:1997bh} in the 37 space. That is, $F'_i$ D5' branes can split
on an NS5 brane (just as a D3 can split on an NS5 brane). In this system, there are two independent $U(F'_i)$ groups (they are gauge groups from the point of view of the D5' worldvolume theory, but global symmetries for the gauge theory living on the D3 branes we consider) associated with the two halves of the D5' brane. In the same way, the mirror NS5' can split on the mirror D5, giving rise to the same enhancement.

We can make use of this enhanced global symmetry in order to introduce Chern-Simons terms in the theory. One way to generate a CS term is to integrate out massive fermions. For every fundamental or anti-fundamental fermion we integrate out, we get a CS term of magnitude 1/2, but with the sign given by the sign of the fermion mass. Adding a mass term $m Q \tilde{Q}$ to the superpotential gives mass to both a fundamental and anti-fundamental fermion, but with opposite sign, thereby generating no net CS term. Another option is to turn on a ``real mass" term. This term can be understood as a spurion. One weakly gauges some global $U(1)$ symmetry, gives a vacuum expectation value to the scalar in this ${\cal N}=2$ vector multiplet, and then sets the gauge coupling to zero. This gives a mass to all fermions proportional to their charge under the global $U(1)$. In the ${\cal N}=4$ theory, we can for example pick the diagonal $U(1)$ of the global $U(F_i)$ flavor symmetry associated with the $F_i$ D5 branes on the $i$-th interval. This, however, will once more fail to produce a CS term as again the fermions in $Q$ and $\tilde{Q}$ have opposite sign mass terms. It is precisely the enhanced $U(F'_i) \times U(F'_i)$ global symmetry of the D5' brane that allows us to produce a CS term. In particular, picking the axial $U(1)$ gives the same sign mass term to both $Q$ and $\tilde{Q}$. A net CS term of level one is generated per flavor. In terms of branes, this means that one pulls apart the intersection of the NS5 and $F'_i=k_{CS}$ D5' branes to generate an
intermediate $(k_{CS},1)$ brane \cite{Bergman:1999na}.
For an NS5 brane separating two intervals, it was argued in \cite{Aharony:2008ug} that this construction gives a
CS term of $+k_{CS}$ to one of the two factors (say the left one)
and $-k_{CS}$ to the other one. To reproduce
this effect from the field theory side, one needs
to account for the phenomenon of flavor doubling \cite{Brunner:1998jr}. While
a single stack of $F_i'$ D5' branes typically gives massless flavors to the $i$-th gauge group factor, once the D5's sit on
top of the NS5 brane they actually give massless fundamental
flavor to both of the neighboring gauge groups. The fundamentals
couple via cubic superpotentials to the bifundamental matter
localized at that NS5 brane. These cubic superpotentials break
the global symmetry from the $U(2F_i)^2$ one would get from two
sets of massless flavors to the $U(F_i)^2$ that is the
correct global symmetry of the theory.
With the brane orientations we are
using here, one obtains an ${\cal N}=2$ CS-matter theory.
The ${\cal N}=3$ theory can be obtained by further
rotating the $(k_{CS,1})$ brane\footnote{\label{N2ABJM}Equivalently, one could have initially
rotated the D5' and NS5 branes. As long as they share five
worldvolume directions, they can form a $(p,q)$ web in the transverse space.
For example, let us look at the ABJM construction. Start with two NS5 branes
and $k_{CS}$ D5' branes. Moving the D5' branes on top of one of the NS5 branes
and splitting them, one obtains an ${\cal N}=2$ supersymmetric gauge theory.
It has the same $U(N) \times U(N)$ gauge group as ABJM with CS terms
of level $k_{CS}$ and $-k_{CS}$ respectively. It has, however, a massless
adjoint
chiral multiplet in each gauge group. Correspondingly, even after the CS
terms are introduced from splitting the D5' branes, one still has a Coulomb
branch associated with vacuum expectation values for $X$. In the brane
language, this branch captures the motion of the D3 branes along the common
45 directions of the branes. To get the ${\cal N}=3$ supersymmetric ABJM theory
(which gets enhanced to ${\cal N}=6$ in the IR), one needs to first rotate
the D5' and one of the NS5 branes together by an angle
$\arctan(k_{CS})$. This rotation introduces a mass term for the two
adjoint chirals proportional to $k_{CS}$. This
is related by ${\cal N}=3$ supersymmetry to the mass the
gauge boson pick up due to the CS term. After integrating
out these massive adjoints, one obtains the characteristic quartic
superpotentials of ABJM.}.

\begin{FIGURE}
{
    \label{genmirror}
    \centerline{\includegraphics[scale=.53]{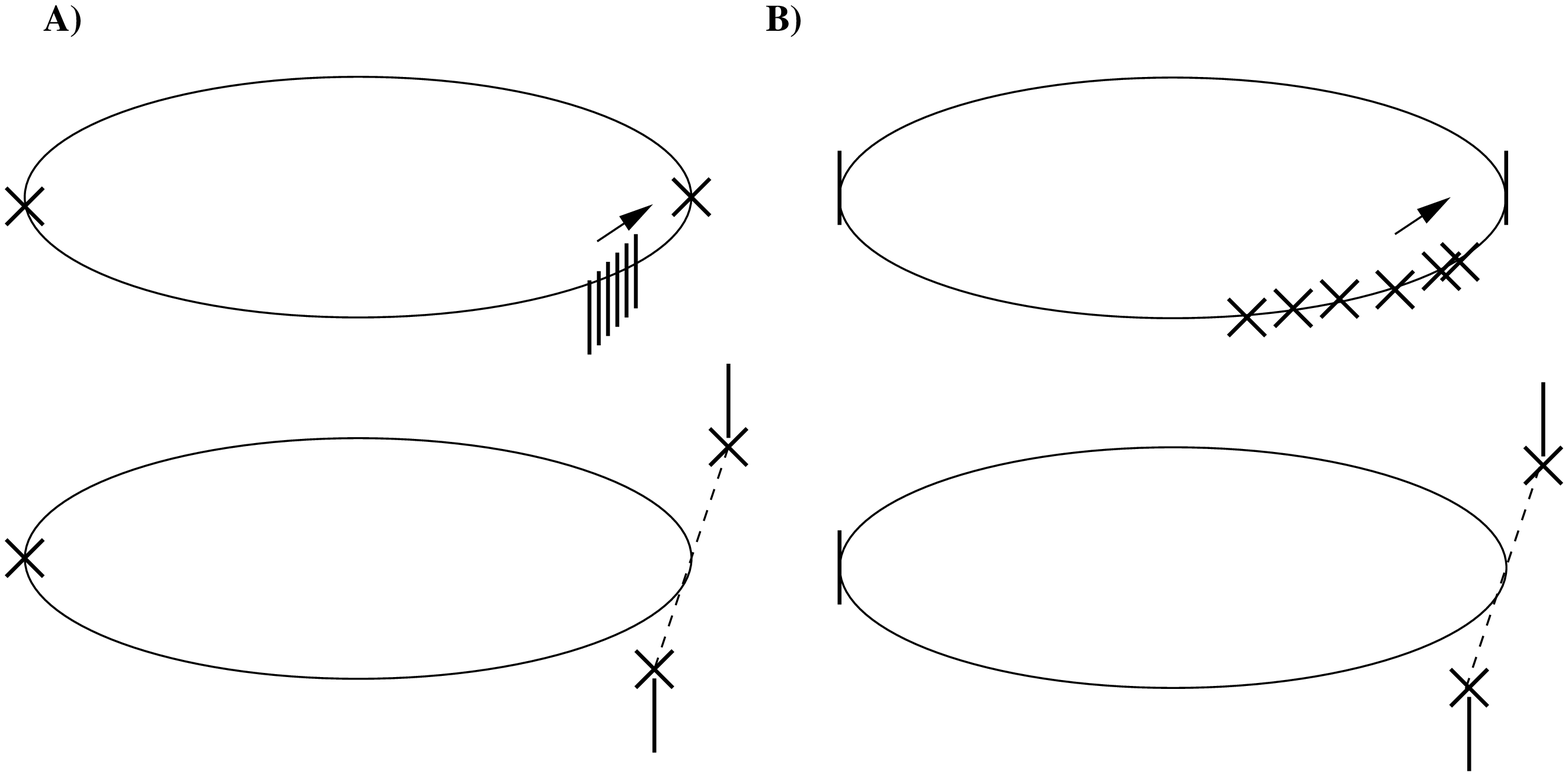}}
    \caption
    {
The mirror of a theory with a $(k_{CS},1)$ brane. In the electric theory
of panel A) the theory with the $(k_{CS},1)$ brane can be obtained
by a deformation of a theory with NS5 and D5' branes only (the former are again
depicted by crosses; this time all vertical lines are D5' branes). In this way,
the mirror theory in panel B) with an $(1,k_{CS})$ brane can also
be obtained as a deformation of a setup with NS5' and D5 branes only, as
described in the text.
     }
}
 \end{FIGURE}

The process of adding a ``real mass" term to the electric theory is depicted in panel A of Figure (\ref{genmirror}). In the S-dual mirror configuration, the same process is implemented by splitting a stack of NS5' branes on a D5 brane into a single $(1,k_{CS})$ brane, as depicted in panel B of (\ref{genmirror}).  From this, we can give an operational definition of  the mirror of (say) the ABJM theory with level $k_{CS}$.
Let us demonstrate this in-principle prescription with a simple example.
Take the mirror pair
displayed in Figure (\ref{genmirror}). For the electric
theory in panel A,
we start with an ${\cal N}=4$ supersymmetric $U(N) \times U(N)$
gauge theory with two bifundamental hypermultiplets. We add $k_{CS}$
${\cal N}=2$ supersymmetry preserving flavors, that is, $k_{CS}$ chiral
multiplets $Q$ and $\tilde{Q}$ in the fundamental and anti-fundamental
representation of both of the $U(N)$ factors (again,
taking into account the flavor doubling effect of \cite{Brunner:1998jr}
). Saying that they are ${\cal N}=2$ flavors means that they can have
no superpotential coupling to the adjoint scalars, as we choose here. The
global symmetry associated with the flavors is therefore $U(k_{CS}) \times
U(k_{CS})$. We now gauge the axial $U(1)$ of the flavor symmetry and introduce an expectation
value for the scalar in this background vector multiplet, giving
a mass to all flavors. Integrating them
out, we get CS terms of level $\pm k_{CS}$ for the
respective gauge group factors.  We are hence
left with an ${\cal N}=2$ version
of ABJM (that is, as we explained in Footnote \ref{N2ABJM},
ABJM without the mass term for the adjoint chiral multiplets in the
two $U(N)$ factors).

The original gauge theory has an easily identified mirror displayed
in panel B. To find the mirror of the resulting ${\cal N}=2$ ABJM
theory, we only need to redo the same steps we took in the electric theory. In the mirror theory, we
start with a $U(N)^{k_{CS}}$ ${\cal N}=4$ gauge theory and
bifundamental matter. There are also two ${\cal N}=2$ flavors
in one of the gauge group factors. There is a manifest $U(2) \times U(2)$ global flavor symmetry. However,
mirror symmetry tells us that deep in the IR of this gauge theory
the full $U(k_{CS}) \times
U(k_{CS}) \times U(2) \times U(2)$ symmetry of the interacting CFT
is realized. So, at the fixed point we can again weakly gauge the
axial combination of the two $U(1)$s in $U(k_{CS}) \times
U(k_{CS})$ and give the background scalar a vacuum expectation
value. The IR fixed point of this field theory is that which one would want to associate
with the brane configuration on the lower right hand side of Figure \ref{genmirror} which
corresponds to the $k_{CS}$ NS5' branes having split on a D5 to form a $
(1,k_{CS})$
brane.

In order to find examples of mirror symmetry that relate
two different field theories with a standard Lagrangian
description, we need to restrict
ourselves to examples with no $(p,q)$ branes with $q > 1$ in either the electric theory or its mirror. But, allowing for $(p,q)$ branes introduces a new freedom in that we are no longer restricted to just using the S-generator of $SL(2,\mathbb{Z})$. Several new examples arise when we also use the T generator. In what follows, we demonstrate that we can find a large number of such examples.

One new situation we will encounter involves
D3 branes stretching between a $(k'_{CS},1)$ brane and a $(k_{CS},1)$ brane.
From our discussion above, this theory will allow a
Lagrangian field theory
interpretation. It can be obtained by starting
with $k_{CS}+k_{CS}'$ D5' branes on an interval between two NS5 branes. Splitting $k_{CS}$ D5's on one NS5 contributes a CS term of $-k_{CS}$, while splitting the remaining $k_{CS}'$ D5's on the other NS5 gives the standard $+k_{CS}'$, so that the net CS term generated has level $k_{CS}' - k_{CS}$. The gauge theory living on the D3 branes between a $(k'_{CS},1)$ brane and a $(k_{CS},1)$ brane is therefore described by a CS theory with level $k_{CS}' - k_{CS}$ (this has also been derived in \cite{Amariti:2009rb}). For a single $(k_{CS},1)$ brane on a circle, we do not expect any CS term as, according to the rules above, the two contributions the $(k_{CS},1)$ brane gives to the left and right gauge group (which now are one and the same) cancel. This is also consistent with the fact that, without any NS5 branes present, one can always perform an $SL(2,\mathbb{Z})$ transformation that brings a $(k_{CS},1)$ brane back to a $(0,1)$ brane (that is an NS5 brane) while leaving all D5 branes untouched.

\subsection{New mirrors with (1,1) branes}

The simplest mirror with (1,1) branes is the original ABJM model at level 1,
that is a $U(N)_{1} \times U(N)_{-1}$ gauge theory
(following the conventions in the literature, we will from
now one use subscripts
to denote the CS level) with two bifundamental hypermultiplets.
The orientation of the (1,1) brane is chosen to preserve
${\cal N}=3$ supersymmetry. The full supersymmetry of the CFT one
obtains in the IR of this configuration is ${\cal N}=8$. The S-dual
configuration has a single D5 brane and a (1,1) brane.
The corresponding gauge theory has no CS term and is simply
$\mathcal N=8$ $U(N)$ super Yang-Mills with a single flavor, as already
pointed out in \cite{Aharony:2008ug}. The IR fixed point of this
theory is also equivalent
to pure ${\cal N}=8$ super Yang-Mills
as well as the Bagger-Lambert theory \cite{Bagger:2007jr}.

\begin{FIGURE}
{
    \label{k1mirrors}
    \centerline{\includegraphics[scale=.53]{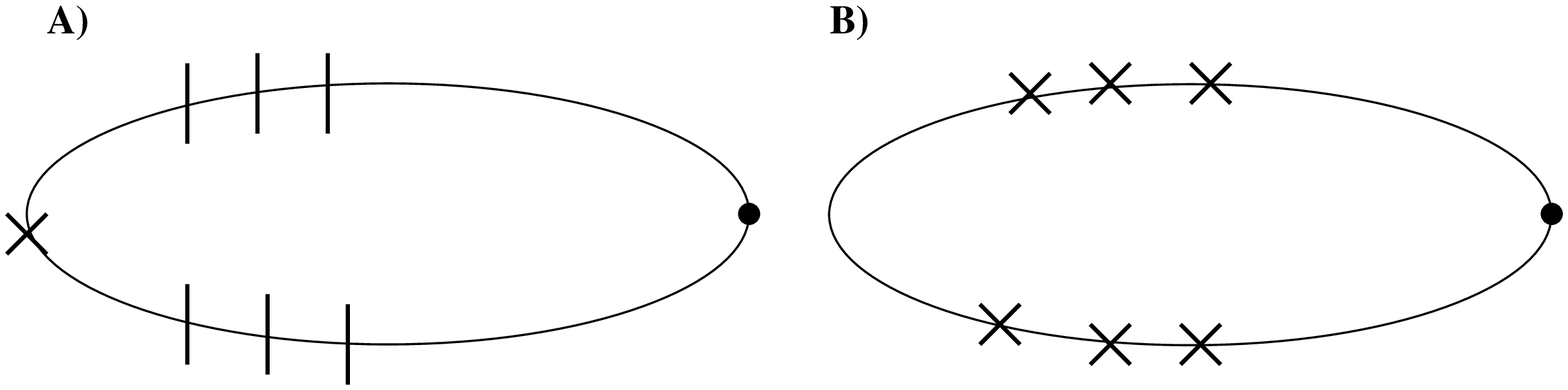}}
    \caption
    {
Standard mirror symmetry (from the S-generator) for brane configurations
involving a (1,1) brane. Crosses are NS5 branes, vertical
lines are D5 branes, and the filled circle represents a (1,1) brane.
Panel A) describes the electric theory and panel B) its mirror.
     }
}
 \end{FIGURE}

We can obviously generate infinitely many new examples along these lines
with arbitrary arrangements of NS5, NS5', D5, D5' and (1,1) branes
around the circle.
Let us demonstrate the power of this construction with one
example, displayed in Figure (\ref{k1mirrors}). For simplicity
we consider a brane setup without NS5' or D5' branes.
For the (1,1) brane, we take the ${\cal N}=2$ preserving
orientation, that is the (1,1) brane was built by splitting a D5' on an NS5
brane.
The electric
theory displayed in panel A is thus a $U(N)_1 \times U(N)_{-1}$
${\cal N}=2$ gauge theory with an adjoint hyper for each of the two $U(N)$ factors and two bifundamental hypers.
In addition each gauge group factor has three fundamental hypers which
couple to the adjoint chirals of each gauge group factor via the standard $\mathcal{N}=4$ cubic superpotential.
The global flavor symmetry is $U(3) \times U(3)$. The mirror gauge
theory has a $U(N)^5 \times U(N)_{1} \times U(N)_{-1}$ gauge group
with adjoint chirals in each of the gauge factors and a single fundamental hyper in one of the $U(N)$ factors.

\subsection{T-mirrors}
\label{Tmirror}
Once one allows for the presence of general $(p,q)$ branes,
the S-generator of the $SL(2,\mathbb{Z})$ S-duality of IIB string theory
is not the only source of gauge theory dualities. One
can also get new non-trivial mirror pairs by including the T-generator;
we will refer to them as T-mirrors. Recall that under the action of $SL(2,\mathbb{Z})$,
the dilaton/axion pair transforms as
\be \tau \rightarrow \frac{ a \tau + b}{c \tau + d}, \ee
with $a,b,c,d \in \mathbb{Z}$ and $ad-bc=1$. A $(p,q)$ brane (where (1,0) is a D5
and (0,1) a NS5) transforms as
\be
\begin{pmatrix} p \\ q \end{pmatrix} \rightarrow
\begin{pmatrix} a & b \\ c & d \end{pmatrix}
\begin{pmatrix} p \\ q \end{pmatrix}.
\ee
There are two new operations we want to examine. The generator
$T^n$ with $a=1$, $b=n$, $c=0$, and $d=1$, leaves all D5 branes untouched
but turns NS5 branes into $(n,1)$ branes. At first sight, this seems
very promising. If we start with a theory that has no $(p,q)$ branes
with $q \geq 2$, we will stay within this class. Unfortunately
this operation does not generate any new mirror pairs. As we have
shown in the last section, for $N$ D3 branes between a $(k'_{CS},1)$
brane and a $(k_{CS},1)$ brane the CS level for the 3d theory is $k_{CS}' - k_{CS}$.
Shifting both $k_{CS}$ and $k'_{CS}$ by the same integer $n$ gives
a new brane configuration, but does not alter any of
the CS terms and hence gives back the same field theory.

\begin{FIGURE}
{
    \label{jy}
   \centerline{\includegraphics[scale=.53]{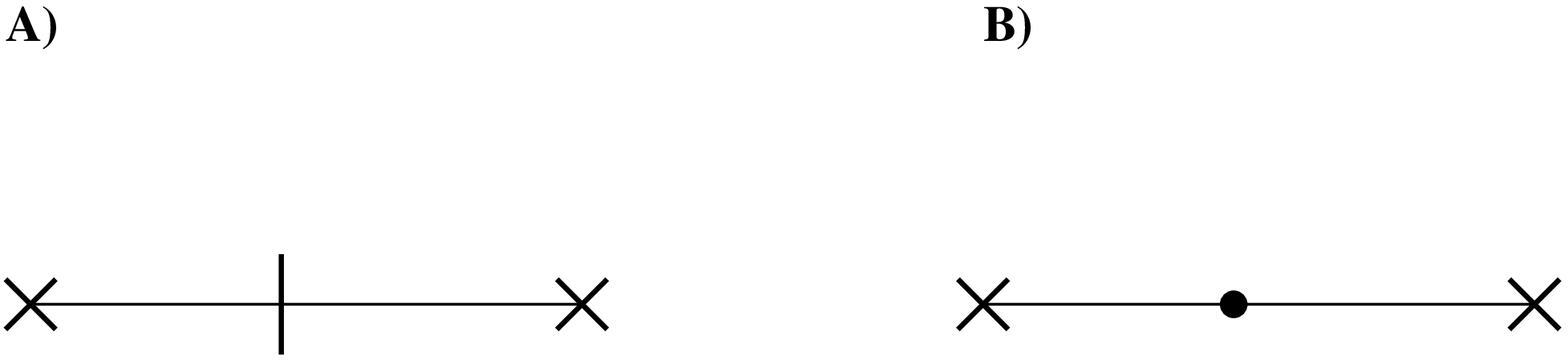}}
    \caption
    {
The T-mirrors of Jafferis and Yin. Crosses are NS5 branes, vertical
lines are D5 branes and the filled circle represents a (1,1) brane.
Panel A) describes the electric theory and panel B) its mirror.
     }
}
 \end{FIGURE}

A more interesting operation is given by $a=c=d=1$ and $b=0$.
This operation leaves the NS5 branes alone but
turns D5 branes into (1,1) branes. While repeated application
of this transformation would generate $(p,q)$ branes with $q \geq 2$,
a single application gives new non-trivial
T-mirror relations between gauge theories with only NS5, D5 and (1,1) branes
(as well as their primed cousins). The first examples of such T-mirrors
have appeared in a paper by Jafferis and Yin \cite{Jafferis:2008em}, studying theories realized by brane configurations on an interval.
For an ${\cal N}=4$ $U(N)$ gauge
theory with $F\leq 3$ flavors (realized as $N$ D3 branes on an interval between
two NS5 branes with $F$ D5 branes on the interval) T-mirrors were
generated
by turning all of the D5 branes into (1,1) branes. The Jafferis-Yin dual
with $F=1$ is displayed in Figure (\ref{jy}). Panel A describes
the electric theory, while panel B gives its T-mirror, a $U(N)_1 \times
U(N)_{-1}$ gauge theory with a single bifundamental hypermultiplet.

In the elliptic models we consider (that is with circular 6 direction),
it is straightforward to generate many more examples of this kind. Take
the generic ${\cal N}=4$ gauge theory we discussed in Section \ref{n4mirror},
that is $k$ NS5 branes with $F_i$ D5 branes on the $i$-th interval.
The electric theory is ${\cal N}=4$ $U(N)^k$ with bifundamental matter
and $F_i$ fundamental hypermultiplets in the $i$-th gauge group factor.
In the T-mirror all D5 branes are turned into (1,1) branes. So
the T-mirror has a $U(N)^{k+F}$ gauge group
with bifundamental matter. Gauge group factors connecting
NS5 branes to (1,1) branes have CS level $\pm 1$ (+ if the (1,1) brane
is on the right, - otherwise) and CS level 0 for gauge group factors connecting
NS to NS or (1,1) to (1,1). All three gauge theories --
the original electric theory, its standard mirror we described
in Section \ref{n4mirror}, and its T-mirror flow to the same interacting
CFT in the IR with a global $\prod_i U(F_i) \times \prod_a U(k_a)$ symmetry.

\subsection{More mirror pairs with $(p,q)$ webs}
\label{webs}

\begin{FIGURE}
{
    \label{junction}
    \centerline{\includegraphics[scale=.53]{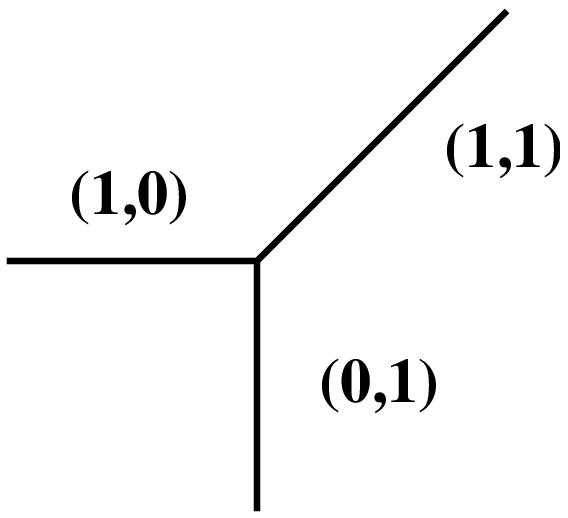}}
    \caption
    {
The junction of an NS5 brane and a D5' brane merging into a (1,1) brane.
     }
}
 \end{FIGURE}

Even more mirror pairs with CS terms can be generated by generalizing our
construction from Section \ref{branesIIB}. So far, we generated CS terms from
a background scalar vev for the vector multiplet associated
with the axial combination of the $U(1)$s in the global $U(F) \times U(F)$
symmetry associated with $F$ D5' branes sitting on top of an NS5 brane.
If one chooses to only pick a single $U(1)$ in, say, the first $U(F)$
factor, one generates a half integer CS term. In terms of
branes, the D3s no longer end on a (1,1) brane but rather at the junction
point where an NS5 and a D5' merge into a (1,1) brane as depicted in
Figure (\ref{junction}). If one takes for example
an elliptic setup with a single NS5 and such a junction, the corresponding
gauge theory is a $U(N)_{1/2} \times U(N)_{-1/2}$ gauge theory with a single fundamental
chiral multiplet in the first $U(N)$ factor and an anti-fundamental in the second $U(N)$ factor.
There is also the standard bifundamental and massless adjoint matter and a cubic
superpotential coupling the flavors to one of the bifundamentals associated with the junction.
This matter content would be anomalous in 4d. In 3d
the theory has a parity anomaly and is in fact only consistent
with half-integer CS level, see e.g. \cite{Aharony:1997bx}.
Were we to turn on a real mass for the
remaining chirals, we would get back the $U(N)_1 \times U(N)_{-1}$ $\mathcal{N}=2$ version of ABJM we
discussed before in Footnote \ref{N2ABJM}.

Our definition of S-duality has so far been accompanied by a relabeling of the 345 and 789 directions. With this relabeling, D5 branes transform into NS5 branes and vice versa, as do D5' branes and NS5' branes. A junction with an NS5 and a D5' brane therefore maps into an analogous junction with an NS5' and a D5' brane. To avoid potential confusion, we will find it convenient to use a different relabeling for the rest of this work, namely one that only exchanges the 3 and 7 directions. Junctions map into themselves under this definition of S-duality, as NS5 branes map to D5's while D5s map to NS5's. 

With this new definition, the mirror to
the theory we just described has a junction and a single D5' on a circle.
The corresponding gauge theory in the IR is simply ${\cal N}=4$ $U(N)$ SYM with three flavors, an adjoint hyper, and cubic superpotential. Again, many more new mirror pairs can be constructed in this way. This construction has already appeared in a paper
by Dorey and Tong \cite{Dorey:1999rb}, where some examples were checked explicitly (that is, they verified that the moduli spaces of electric and mirror theory match).

\section{There is only one M-theory: geometric duality and AdS/CFT}
\label{oneM}
\subsection{M-theory uplift}

All of our examples of mirror symmetric pairs with ${\cal N}=2$ and ${\cal N}=3$
represent two (or more) distinct gauge theories that flow to the same fixed point
in the IR.
As in the ${\cal N}=4$ examples we discussed in Section \ref{n4mirror},
the various brane configurations in IIB simply lift to M2 branes at the singularity of one and the same Calabi-Yau four-fold in M-theory.

The generic ${\cal N}=2$ theory we can realize with our IIB brane setups
has NS5, NS5', D5 and D5' branes or their cousins at $\mathcal{N}=2$
preserving rotation angles. For a special subclass of these setups, we
can easily find the associated M-theory lift. At large $N$, this also yields an explicit
AdS/CFT dual to the field theory at the fixed point. Consider a network
of NS5, D5' and $(p,q)$ webs formed from the two objects as in Section \ref{webs}. Deep in the IR,
all of these 5-branes intersect and the whole collection of NS5, D5', and $(p,q)$
webs can be replaced with a single web.

\begin{FIGURE}
{
    \label{spptoric}
    \centerline{\includegraphics[scale=.53]{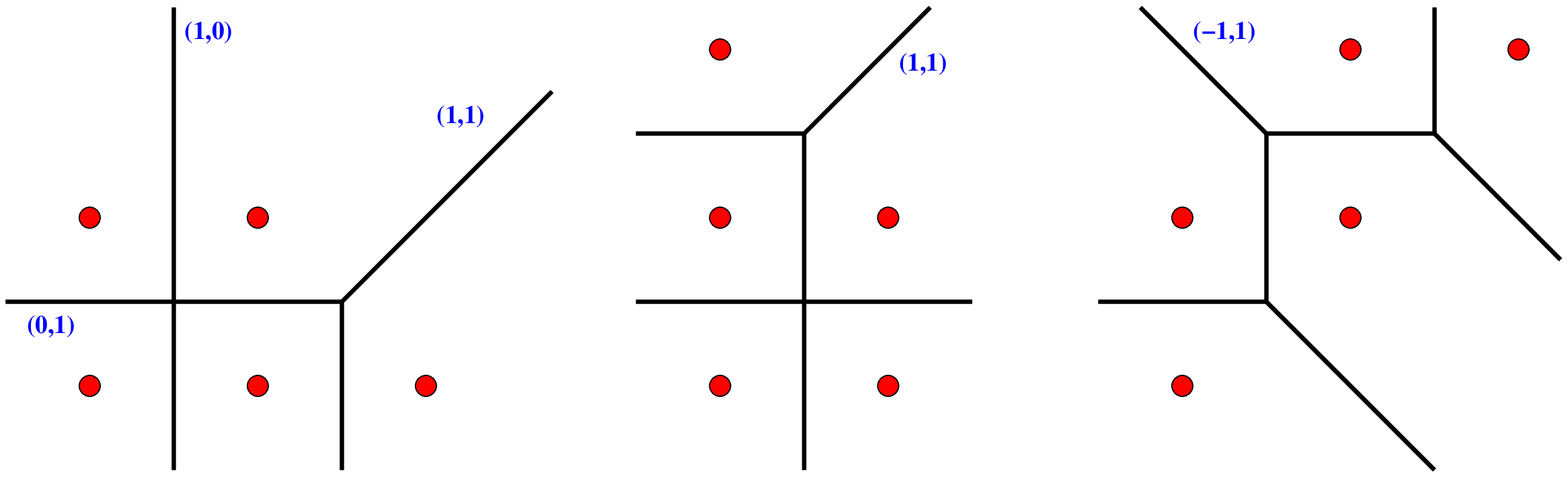}}
    \caption
    {
The
toric diagram of the suspended pinch point singularity together
with
its dual $(p,q)$ web. Also shown are two different representations of
the toric diagram/dual web related to the leftmost diagram
by toric duality/$SL(2,\mathbb{Z})$ S-duality.
     }
}
 \end{FIGURE}

In this case we can use the result of Leung and Vafa \cite{Leung:1997tw} who showed that the generic $(p,q)$
web directly lifts to a toric CY 3-fold in M-theory. The graph of the $(p,q)$
web and the standard representation of the toric diagram are dual to each other
in the sense of exchanging faces and vertices.
As an example, the toric diagram for the suspended pinch
point singularity and its dual $(p,q)$ web are depicted in
the leftmost panel of Figure
(\ref{spptoric}). In this way, we can easily associate a toric $CY_3$ to any
gauge theory we obtain from putting NS5 and D5' branes as well
as $(p,q)$ webs made out of these objects. The
gauge theory (as well as its mirror) flows to the CFT described
by M2 branes probing this 3-fold.

In general, an $\mathcal{N}=2$ theory on M2 branes can be realized by a stack of M2s probing a toric $CY_4.$
We therefore see that the restriction to NS5 and D5' branes (as well as their associated webs) in our IIB construction corresponds
to specializing the 4-fold we generically expect to be of the form $CY_3\times \mathbb{R}^2$. In terms
of supersymmetry, ${\cal N}=2$ is preserved in either case: for the 3-fold,
the probe M2 breaks half of the supersymmetry preserved by
the 3-fold. For the 4-fold the M2 brane can be added without any
further SUSY breaking. This is also clear from the brane setup.
The M-theory description also provides us with an explicit AdS/CFT realization
of the IR fixed point when $N \gg 1$. Toric CY 3-folds
can be written as a cone over some 5d Einstein-Sasaki base $B$, with metric
\be
ds^2_{3-fold} = d\rho^2 + \rho^2 ds^2_B.
\ee
Placing $N$ M2 branes at the tip of the cone, one gets a near horizon
geometry of the form
\begin{eqnarray}
\nonumber
ds^2 &=& \frac{r^2 + \rho^2}{L^2} (- dt^2 + d \vec{x}^2) +
\frac{L^2}{r^2 + \rho^2} ( dr^2 + r^2 d \alpha^2 + d \rho^2 + \rho^2 ds^2_B)\\
&=& \frac{R^2}{L^2} (- dt^2 + d \vec{x}^2) +
\frac{L^2}{R^2} dR^2 + L^2 \left ( d \theta^2 + \cos^2\theta  \,d \alpha^2
+ \sin^2\theta  \, ds^2_B \right ).
\end{eqnarray}
where $t$ and $\vec{x}$ are the three directions along the M2 worldvolume,
$r$ and $\alpha$ are spherical coordinates on $\mathbb{R}^2$,
and in the second line we introduced $R^2 = r^2 + \rho^2$ with
$r= \cos \theta \, R$ and $\rho = \sin \theta \, R$.
This is a solution to 11d supergravity with $N$ units of 7-form
flux through the 7d internal space.

\subsection{Geometric duality}
\label{geoDuality}
M2 branes on a toric $CY_3$ have an alternative description.
Instead of going directly down to IIB using the toroidal action of the toric geometry,
we can compactify and reduce on a trivial circle in the $\mathbb{R}^2$ transverse to both the M2s and the $CY_3$.
In this way, we obtain D2 branes in IIA string theory on the same 3-fold. While the former procedure gives the
CS theories we described before (as well as their mirror symmetric partners),
the latter gives yet another completely different description of a gauge
theory with the same IR fixed point. We will refer to this equivalence between these theories as ``geometric duality." To read off this new gauge theory of D2s at the singularity of the $CY_3$,
one can bring all the powerful techniques developed in recent years to bear. There is a well developed algorithm
\cite{Beasley:1999uz,Feng:2000mi} following the basic idea of the resolution of orbifold singularities suggested in
\cite{Morrison:1998cs} that associates a quiver gauge theory
with any D-brane on a toric $CY_3$ singularity. While
this technique has been mostly exploited in the context of D3 branes on a $CY_3$ singularity, exactly the
same gauge group describes D2 branes on the same $CY_3$. The resulting gauge theory
can be read off in the usual way from a ``brane tiling" diagram \cite{Franco:2005rj,Franco:2005sm}. No CS terms are involved in this description. Sometimes a further T-duality can be performed to a brane setup in type IIB
involving only NS5 and NS5' branes and no D5 branes of any kind \cite{Uranga:1998vf,Dasgupta:1998su}.

\begin{FIGURE}
{
    \label{spp}
    \centerline{\includegraphics[scale=.53]{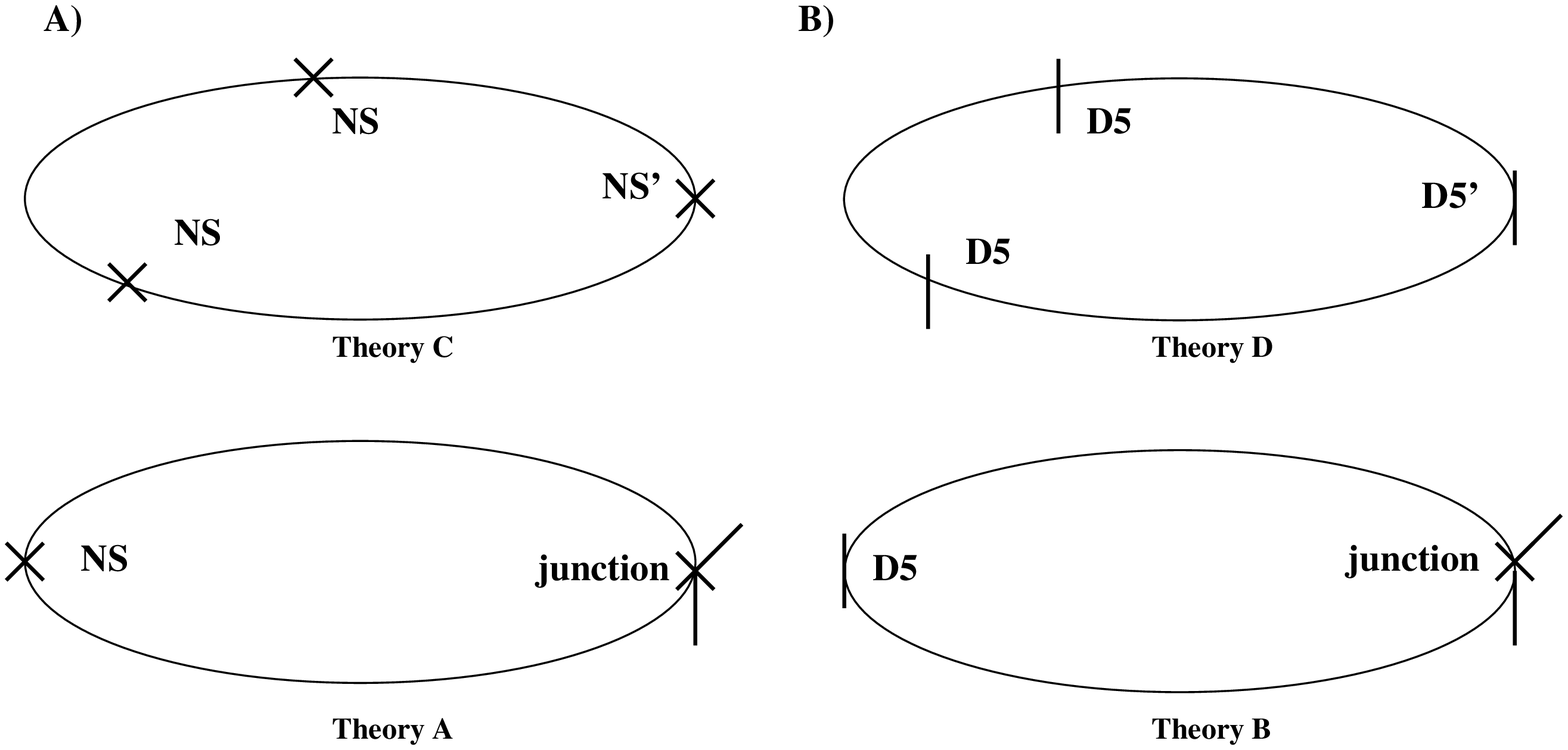}}
    \caption
    {
Four different realization of a gauge theory that flows to the
theory of M2 branes probing the suspended pinch point
singularity in M-theory. The top row has the standard
IIB brane setups T-dual to D2 branes on that singularity (panel A) and its mirror
(panel B).
The second row is the "geometric dual" of the first setup (panel A) and
its mirror (panel B). The theories in the second row involve the
basic junction of Figure (\ref{junction}), but only the theory
in the second row of panel A has non-trivial CS terms. As we describe
in the text, the two brane setups in panel B actually give
identical gauge theories, and so there are only three distinct gauge
groups encoded in the four brane configurations.
}
}
 \end{FIGURE}

As an example, let us look once more at the suspended pinch point.
As depicted in the leftmost panel of
Figure (\ref{spptoric}), the corresponding $(p,q)$
web has one NS5 brane and one junction merging an NS5 and a D5' into a (1,1)
brane. Suspending $N$ D3 branes between these
as displayed in the lower diagram of panel A in Figure (\ref{spp}),
we get exactly the electric
theory described in the at the end of Section \ref{webs}, where the NS5 sits on top of the junction. The corresponding
quiver diagram, as well as the quivers for the dual
theories we are about to discuss, are
displayed in Figure (\ref{sppquivers}).  This gauge theory can be summarized as

\noindent{\bf Theory A:} a $U(N)_{1/2} \times
U(N)_{-1/2}$ gauge theory with a single (anti-) fundamental
chiral multiplet in the (second) first gauge group factor as well
as the standard bifundamentals, and adjoints for
each gauge group. There is a cubic superpotential coupling the
flavors to one of the bifundamentals.

\noindent The suspended pinch point arises in this case as the quantum corrected Coulomb branch
of this theory. This is similar to the other examples of toric geometries on 3d Coulomb
branches of
\cite{Dorey:1999rb}. The mirror dual of this configuration (lower diagram of panel B in Figure
(\ref{spp}) ) has the NS5 brane replaced by a D5', which sits on top of the junction. This is

\noindent{ \bf Theory B:}
an $\mathcal{N}=4$ $U(N)$ gauge theory with no CS term, but three fundamental
hypermultiplets, an adjoint hyper, and cubic superpotential.

\noindent One hypermultiplet comes from the (flavor doubled) half D5' in the junction, two more from
the D5' (again accounting for flavor doubling).

The geometric dual of Theory A is the standard quiver gauge theory of the suspended pinch point,

\noindent {\bf Theory C:} a $U(N)^3$ quiver gauge theory with bifundamentals connecting neighboring
gauge groups and an extra adjoint in one of the gauge group factors.

\noindent The IIA setup with D2s probing the suspended pinch point singularity as in Theory C can
also be nicely captured by a T-dual setup \cite{Uranga:1998vf}
with two NS5 branes and one NS5' as
depicted in the upper diagram of panel A in Figure (\ref{spp}). This also
makes it easy to read off, via S-duality, the mirror description of Theory C (displayed in the upper diagram
of panel B in Figure (\ref{spp})).  The mirror is

\noindent {\bf Theory D:} a ${\cal N}=8$ $U(N)$ gauge with theory with three fundamental hypermultiplets and a superpotential coupling
$ W = X Q_i \tilde{Q}_i + Y T \tilde{T} $
where ($\tilde{Q}_i$) $Q_i$ (with $i=1,2$) and ($\tilde{T}$) $T$ are the (anti-) fundamental chiral multiplets in the three hypers and $X$
and $Y$ are two different adjoint chiral multiplets from the ${\cal N}=8$ vector multiplet.

In this example, Theory D is actually identical to Theory B. The matter content is clearly identical and the cubic superpotential of the flavor doubling mechanism \cite{Brunner:1998jr} at work in Theory B gives back the superpotential of Theory D. This allows us to ``prove" this whole duality cycle by simply using the usual mirror symmetry from S-duality of brane setups: C is mirror to D, D is identical to B, and B is mirror to A. For more
complicated toric diagrams, the four theories A, B, C, and D will all be different. Theories A and B generically have CS terms or even $(p,q)$ branes with $q \geq 2$. Theory C is typically a quiver gauge theory without fundamentals, while Theory D is usually ${\cal N}=8$ $U(N)$ gauge
theory coupled to fundamental matter with a non-trivial superpotential. They all flow to the same IR fixed point.

Let us also briefly mention the example of the conifold, as this is typically the first example worked out. Here, the labels Theory A, B, C, or D will refer to the corresponding theory for the case of the conifold. The brane configuration for Theory A has one NS5 and one D5' in a 'cross' configuration, and (by S-duality) so does theory B. The corresponding gauge group is a single ${\cal N}=8$ gauge theory with 2 fundamentals (accounting for flavor doubling) coupling to two different adjoints. Theory C has an NS5 brane and an NS5' brane, so it is a $U(N) \times U(N)$ quiver gauge theory with bifundamentals.  This is the standard Klebanov-Witten theory, but in 3d \cite{Klebanov:1998hh}. Theory D has a D5 and a D5', which again gives the same matter content and superpotential as Theories A and B. So, while in this case we still uncover an interesting duality network between brane configurations (we can either have two NS-type branes, two D-type branes or one of each all describing the same singularity) geometric duality does not provide any new field theory dualities\footnote{One potential objection
is that the theory with a single NS5 and a single D5' can't possibly be equivalent to the other two: under T-duality, a single NS5 brane disappears, so the dual theory seems to be $N$ D2 branes transverse to a single D6 brane. As shown in \cite{Park:1999eb}, the T-duality of NS5 branes into an orbifold geometry in the presence of D5' branes is more subtle, and indeed a single D5' on top of an NS5 T-dualizes into two distinct, orthogonal D6 branes. This is the IIA manifestation
of flavor doubling. In the IIA setting, the theory coming from one NS5 and one D5' brane is then manifestly the same as the one coming from just a D5/D5' pair.}. We can however generate a new mirror to the conifold using our results in Section \ref{Tmirror}. In particular, recall that there is an $SL(2,\mathbb{Z})$ transformation that leaves NS5 branes untouched but turns D5s into (1,1) branes. This transformation takes Theory A/B for the conifold into a brane setup with an NS5 brane on top of a (1,1) brane.  This brane setup gives us a different mirror, namely $U(N)_{1}\times U(N)_{-1}$ gauge theory with massless adjoints and some quartic superpotential. Note that this setup is the same as our $\mathcal{N}=2$ version of ABJM in Footnote \ref{N2ABJM} at level 1, in the limit that the NS5 sits on top of the (1,1) brane. This result was already obtained in \cite{Aganagic:2009zk}.

So far, we have only considered theories corresponding
to the special case of M2 branes probing non-compact $CY_3$ singularities.
It would be interesting to work
out the detailed map between a generic IIB brane configurations and the geometry
of the full 4-fold that one gets in M-theory.
Several proposals have been made for identifying which
gauge group should go with which 4-fold singularity
\cite{Martelli:2008si,Hanany:2008cd,
Hanany:2008fj,Jafferis:2008qz,Ueda:2008hx,Hanany:2008gx,Imamura:2008qs,Lee:2007kv}. However, it is yet unclear
if and how these proposals can be related to a D-brane construction.

An important step towards that goal
has recently been made in \cite{Aganagic:2009zk}.
Aganagic showed that an M2 brane on
a 4-fold can reduce to a D2 brane on a 3-fold fibered over a line with RR
flux. This RR flux gives rise to CS terms. If one repeats the same
steps for a 3-fold in M-theory, one gets a 2-fold fibered
over a line with RR flux in IIA. Presumably, this gives a IIA realization
of the theories we get from directly reducing to IIB from M-theory
using the Leung-Vafa construction \cite{Leung:1997tw}.

\begin{FIGURE}
{
    \label{sppquivers}
    \centerline{\includegraphics[scale=.8]{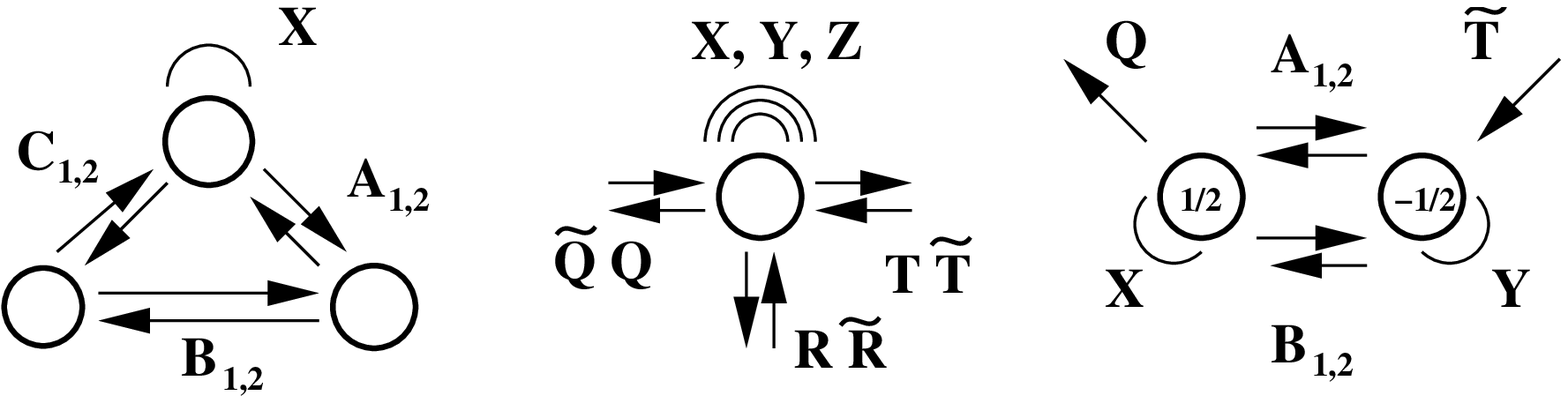}}
\vspace{-29pt}
    \caption
    {
Quiver diagrams for the different gauge theories associated
with the suspended pinch point singularity as described in the text.
We use $\mathcal{N}=2$ notation to represent the different multiplets in each quiver.
From left to right, we have Theory C, Theory B = Theory D and Theory A
respectively. Nodes are $U(N)$ gauge groups (CS terms, when present, are
written inside the node). Every line is a chiral multiplet. The
superpotentials are $W=X (A_1 A_2 - C_1 C_2) + B_1 B_2 C_2 C_1
- B_2 B_1 A_2 A_1$  , $W = X (Q \tilde{Q}
+ T \tilde{T}) + Y R\tilde{R}$ and $W= Q A_2 \tilde{T} + X (A_1 A_2 - B_1 B_2)
- Y (A_2 A_1 + B_2 B_1)$ respectively (a trace over gauge indices being
implicit).
     }
}
 \end{FIGURE}

\section{Toric duality is Seiberg duality and toric duality
is mirror symmetry: a duality of dualities}
\label{dualDualities}
We have already identified two different sources of universality
in 3d supersymmetric gauge theories. The first set comes from mirror symmetries
generated by the full $SL(2,\mathbb{Z})$ S-duality of IIB string
theory, which can generate large families of different gauge
theories that all flow to the same IR fixed point. The geometric duality
described in the Section \ref{geoDuality} also allows us to identify to different families
with the same IR fixed point. There is yet another well studied equivalence
of gauge theories realized on brane probes of toric singularities: toric
duality \cite{Feng:2000mi,Feng:2001xr}. The toric diagram for a 3 (complex)
dimensional toric geometry is given by vectors in a 3d lattice. The lattice has
an $SL(3,\mathbb{Z})$ symmetry, under which the various vectors that describe the
geometry transform, but the toric manifold they describe is the same. For that toric manifold to be Calabi-Yau,
the endpoints of all the vectors defining the toric diagram have to live in a
plane (and it is only this plane that is displayed in the typical toric
diagrams for non-compact CYs). Thus, only an $SL(2,\mathbb{Z})$ subgroup of the full $SL(3,\mathbb{Z})$ acts
on the toric diagram in a non-trivial way, giving different diagrams associated with the same $CY_3$. Since we can obtain
a $(p,q)$ web associated with each description of that $CY_3$, there are many different IIB brane constructions that all lift to
M2 branes probing the same $CY_3$. There are therefore a class of theories that clearly flow
to the same CFT in the IR.

Moreover, if we use the Leung Vafa procedure to directly reduce down at $CY_3$ singularity to IIB, the $SL(2,\mathbb{Z})$ relation of toric diagrams  just becomes the $SL(2,\mathbb{Z})$ S-duality of IIB. We identified this IIB S-duality before with mirror symmetry in the gauge theory, so it is clear that in this picture toric duality is nothing but mirror symmetry.
For example, for the suspended pinch point
discussed in Section \ref{geoDuality} we can associate
different $(p,q)$ webs with the different but equivalent toric diagrams of
Figure (\ref{spptoric}). The associated gauge theories are related by mirror
symmetry.
In Section \ref{geoDuality} we only made use of the first two
realizations, which are related to each other via S-duality.
The third diagram would correspond to one of the T-mirrors we
can generate for this gauge theory.
All these theories flow to the same IR fixed point, namely M2s probing the suspended pinch point singularity.

The story is different however in the picture we get from reducing to D2 branes in type IIA. As we mentioned at the end of Section \ref{geoDuality}, for D2 branes on the 3-fold the analysis is a straightforward dimensional reduction of the results of
\cite{Feng:2000mi,Feng:2001xr} for D3 branes on the same singularity. In this setting, toric duality is actually identified in \cite{Feng:2001bn,Beasley:2001zp} with Seiberg duality \cite{Seiberg:1994pq}. Seiberg duality survives
the reduction from 4d to 3d virtually unchanged as first demonstrated in \cite{Karch:1997ux,Aharony:1997gp}, so the identification of toric duality with Seiberg duality will still be valid\footnote{Seiberg duality in 3d has shown
to also generalize to theories with CS terms in \cite{Giveon:2008zn} and was argued to still capture toric duality in the general case on a CY 4-fold in \cite{Amariti:2009rb}. For the 3-fold case we are discussing, the situation is much simpler as we only need to apply Seiberg duality without CS terms.}. As the two different reductions give rise to two different realizations of the same CFT and hence are related by our new geometric duality, we find that under geometric duality mirror symmetry
(the realization of toric duality in the IIB picture)
gets mapped to Seiberg duality (the realization of toric duality in the IIA picture).
For the generic case of the 4-fold,
it is probably not even meaningful anymore to distinguish between mirror symmetry, Seiberg duality and geometric duality (for a recent discussion of Seiberg duality and toric duality in this context see e.g. \cite{Amariti:2009rb,Franco:2009sp,Davey:2009sr}).
They are all different manifestations of the fact that many D-brane configurations in type II string theory map to one and the same geometry in M-theory.

\section*{Acknowledgements}
We would like to thank D.~Tong and E.~Witten for useful email correspondence.
This work was supported in part by the U.S. Department
    of Energy under Grant No.~DE-FG02-96ER40956.

\bibliography{abjm}

\providecommand{\href}[2]{#2}\begingroup\raggedright\begin{thebibliography}{10}

\bibitem{Intriligator:1996ex}
K.~A. Intriligator and N.~Seiberg, {\it {Mirror symmetry in three dimensional
  gauge theories}},  {\em Phys. Lett.} {\bf B387} (1996) 513--519,
  \href{http://xxx.lanl.gov/abs/hep-th/9607207}{{\tt hep-th/9607207}}.

\bibitem{Kapustin:1999ha}
A.~Kapustin and M.~J. Strassler, {\it {On Mirror Symmetry in Three Dimensional
  Abelian Gauge Theories}},  {\em JHEP} {\bf 04} (1999) 021,
  \href{http://xxx.lanl.gov/abs/hep-th/9902033}{{\tt hep-th/9902033}}.

\bibitem{Sachdev:2008wv}
S.~Sachdev and X.~Yin, {\it {Deconfined criticality and supersymmetry}},
  \href{http://xxx.lanl.gov/abs/0808.0191}{{\tt 0808.0191}}.

\bibitem{Aharony:1997bx}
O.~Aharony, A.~Hanany, K.~A. Intriligator, N.~Seiberg, and M.~J. Strassler,
  {\it {Aspects of N = 2 supersymmetric gauge theories in three dimensions}},
  {\em Nucl. Phys.} {\bf B499} (1997) 67--99,
  \href{http://xxx.lanl.gov/abs/hep-th/9703110}{{\tt hep-th/9703110}}.

\bibitem{Aharony:2008ug}
O.~Aharony, O.~Bergman, D.~L. Jafferis, and J.~Maldacena, {\it {N=6
  superconformal Chern-Simons-matter theories, M2-branes and their gravity
  duals}},  {\em JHEP} {\bf 10} (2008) 091,
  \href{http://xxx.lanl.gov/abs/0806.1218}{{\tt 0806.1218}}.

\bibitem{Tong:2000ky}
D.~Tong, {\it {Dynamics of N = 2 supersymmetric Chern-Simons theories}},  {\em
  JHEP} {\bf 07} (2000) 019, \href{http://xxx.lanl.gov/abs/hep-th/0005186}{{\tt
  hep-th/0005186}}.

\bibitem{Aganagic:2009zk}
M.~Aganagic, {\it {A Stringy Origin of M2 Brane Chern-Simons Theories}},
  \href{http://xxx.lanl.gov/abs/0905.3415}{{\tt 0905.3415}}.

\bibitem{Hanany:1996ie}
A.~Hanany and E.~Witten, {\it {Type IIB superstrings, BPS monopoles, and three-
  dimensional gauge dynamics}},  {\em Nucl. Phys.} {\bf B492} (1997) 152--190,
  \href{http://xxx.lanl.gov/abs/hep-th/9611230}{{\tt hep-th/9611230}}.

\bibitem{Borokhov:2002cg}
V.~Borokhov, A.~Kapustin, and X.-k. Wu, {\it {Monopole operators and mirror
  symmetry in three dimensions}},  {\em JHEP} {\bf 12} (2002) 044,
  \href{http://xxx.lanl.gov/abs/hep-th/0207074}{{\tt hep-th/0207074}}.

\bibitem{Ooguri:1995wj}
H.~Ooguri and C.~Vafa, {\it {Two-Dimensional Black Hole and Singularities of CY
  Manifolds}},  {\em Nucl. Phys.} {\bf B463} (1996) 55--72,
  \href{http://xxx.lanl.gov/abs/hep-th/9511164}{{\tt hep-th/9511164}}.

\bibitem{Gregory:1997te}
R.~Gregory, J.~A. Harvey, and G.~W. Moore, {\it {Unwinding strings and
  T-duality of Kaluza-Klein and H- monopoles}},  {\em Adv. Theor. Math. Phys.}
  {\bf 1} (1997) 283--297, \href{http://xxx.lanl.gov/abs/hep-th/9708086}{{\tt
  hep-th/9708086}}.

\bibitem{Tong:2002rq}
D.~Tong, {\it {NS5-branes, T-duality and worldsheet instantons}},  {\em JHEP}
  {\bf 07} (2002) 013, \href{http://xxx.lanl.gov/abs/hep-th/0204186}{{\tt
  hep-th/0204186}}.

\bibitem{Porrati:1996xi}
M.~Porrati and A.~Zaffaroni, {\it {M-theory origin of mirror symmetry in three
  dimensional gauge theories}},  {\em Nucl. Phys.} {\bf B490} (1997) 107--120,
  \href{http://xxx.lanl.gov/abs/hep-th/9611201}{{\tt hep-th/9611201}}.

\bibitem{Witten:2009xu}
E.~Witten, {\it {Branes, Instantons, And Taub-NUT Spaces}},
  \href{http://xxx.lanl.gov/abs/0902.0948}{{\tt 0902.0948}}.

\bibitem{Karch:1998yv}
A.~Karch, D.~Lust, and D.~J. Smith, {\it {Equivalence of geometric engineering
  and Hanany-Witten via fractional branes}},  {\em Nucl. Phys.} {\bf B533}
  (1998) 348--372, \href{http://xxx.lanl.gov/abs/hep-th/9803232}{{\tt
  hep-th/9803232}}.

\bibitem{Douglas:1996sw}
M.~R. Douglas and G.~W. Moore, {\it {D-branes, Quivers, and ALE Instantons}},
  \href{http://xxx.lanl.gov/abs/hep-th/9603167}{{\tt hep-th/9603167}}.

\bibitem{Aharony:2008gk}
O.~Aharony, O.~Bergman, and D.~L. Jafferis, {\it {Fractional M2-branes}},  {\em
  JHEP} {\bf 11} (2008) 043, \href{http://xxx.lanl.gov/abs/0807.4924}{{\tt
  0807.4924}}.

\bibitem{Kitao:1998mf}
T.~Kitao, K.~Ohta, and N.~Ohta, {\it {Three-dimensional gauge dynamics from
  brane configurations with (p,q)-fivebrane}},  {\em Nucl. Phys.} {\bf B539}
  (1999) 79--106, \href{http://xxx.lanl.gov/abs/hep-th/9808111}{{\tt
  hep-th/9808111}}.

\bibitem{Bergman:1999na}
O.~Bergman, A.~Hanany, A.~Karch, and B.~Kol, {\it {Branes and supersymmetry
  breaking in 3D gauge theories}},  {\em JHEP} {\bf 10} (1999) 036,
  \href{http://xxx.lanl.gov/abs/hep-th/9908075}{{\tt hep-th/9908075}}.

\bibitem{Uranga:1998vf}
A.~M. Uranga, {\it {Brane Configurations for Branes at Conifolds}},  {\em JHEP}
  {\bf 01} (1999) 022, \href{http://xxx.lanl.gov/abs/hep-th/9811004}{{\tt
  hep-th/9811004}}.

\bibitem{Aharony:1997ju}
O.~Aharony and A.~Hanany, {\it {Branes, superpotentials and superconformal
  fixed points}},  {\em Nucl. Phys.} {\bf B504} (1997) 239--271,
  \href{http://xxx.lanl.gov/abs/hep-th/9704170}{{\tt hep-th/9704170}}.

\bibitem{Aharony:1997bh}
O.~Aharony, A.~Hanany, and B.~Kol, {\it {Webs of (p,q) 5-branes, five
  dimensional field theories and grid diagrams}},  {\em JHEP} {\bf 01} (1998)
  002, \href{http://xxx.lanl.gov/abs/hep-th/9710116}{{\tt hep-th/9710116}}.

\bibitem{Brunner:1998jr}
I.~Brunner, A.~Hanany, A.~Karch, and D.~Lust, {\it {Brane dynamics and chiral
  non-chiral transitions}},  {\em Nucl. Phys.} {\bf B528} (1998) 197--217,
  \href{http://xxx.lanl.gov/abs/hep-th/9801017}{{\tt hep-th/9801017}}.

\bibitem{Amariti:2009rb}
A.~Amariti, D.~Forcella, L.~Girardello, and A.~Mariotti, {\it {3D Seiberg-like
  Dualities and M2 Branes}},  \href{http://xxx.lanl.gov/abs/0903.3222}{{\tt
  0903.3222}}.

\bibitem{Bagger:2007jr}
J.~Bagger and N.~Lambert, {\it {Gauge Symmetry and Supersymmetry of Multiple
  M2-Branes}},  {\em Phys. Rev.} {\bf D77} (2008) 065008,
  \href{http://xxx.lanl.gov/abs/0711.0955}{{\tt 0711.0955}}.

\bibitem{Jafferis:2008em}
D.~L. Jafferis and X.~Yin, {\it {Chern-Simons-Matter Theory and Mirror
  Symmetry}},  \href{http://xxx.lanl.gov/abs/0810.1243}{{\tt 0810.1243}}.

\bibitem{Dorey:1999rb}
N.~Dorey and D.~Tong, {\it {Mirror symmetry and toric geometry in three
  dimensional gauge theories}},  {\em JHEP} {\bf 05} (2000) 018,
  \href{http://xxx.lanl.gov/abs/hep-th/9911094}{{\tt hep-th/9911094}}.

\bibitem{Leung:1997tw}
N.~C. Leung and C.~Vafa, {\it {Branes and toric geometry}},  {\em Adv. Theor.
  Math. Phys.} {\bf 2} (1998) 91--118,
  \href{http://xxx.lanl.gov/abs/hep-th/9711013}{{\tt hep-th/9711013}}.

\bibitem{Beasley:1999uz}
C.~Beasley, B.~R. Greene, C.~I. Lazaroiu, and M.~R. Plesser, {\it {D3-branes on
  partial resolutions of abelian quotient singularities of Calabi-Yau
  threefolds}},  {\em Nucl. Phys.} {\bf B566} (2000) 599--640,
  \href{http://xxx.lanl.gov/abs/hep-th/9907186}{{\tt hep-th/9907186}}.

\bibitem{Feng:2000mi}
B.~Feng, A.~Hanany, and Y.-H. He, {\it {D-brane gauge theories from toric
  singularities and toric duality}},  {\em Nucl. Phys.} {\bf B595} (2001)
  165--200, \href{http://xxx.lanl.gov/abs/hep-th/0003085}{{\tt
  hep-th/0003085}}.

\bibitem{Morrison:1998cs}
D.~R. Morrison and M.~R. Plesser, {\it {Non-spherical horizons. I}},  {\em Adv.
  Theor. Math. Phys.} {\bf 3} (1999) 1--81,
  \href{http://xxx.lanl.gov/abs/hep-th/9810201}{{\tt hep-th/9810201}}.

\bibitem{Franco:2005rj}
S.~Franco, A.~Hanany, K.~D. Kennaway, D.~Vegh, and B.~Wecht, {\it {Brane Dimers
  and Quiver Gauge Theories}},  {\em JHEP} {\bf 01} (2006) 096,
  \href{http://xxx.lanl.gov/abs/hep-th/0504110}{{\tt hep-th/0504110}}.

\bibitem{Franco:2005sm}
S.~Franco {\em et~al.}, {\it {Gauge theories from toric geometry and brane
  tilings}},  {\em JHEP} {\bf 01} (2006) 128,
  \href{http://xxx.lanl.gov/abs/hep-th/0505211}{{\tt hep-th/0505211}}.

\bibitem{Dasgupta:1998su}
K.~Dasgupta and S.~Mukhi, {\it {Brane Constructions, Conifolds and M-Theory}},
  {\em Nucl. Phys.} {\bf B551} (1999) 204--228,
  \href{http://xxx.lanl.gov/abs/hep-th/9811139}{{\tt hep-th/9811139}}.

\bibitem{Klebanov:1998hh}
I.~R. Klebanov and E.~Witten, {\it {Superconformal field theory on threebranes
  at a Calabi-Yau singularity}},  {\em Nucl. Phys.} {\bf B536} (1998) 199--218,
  \href{http://xxx.lanl.gov/abs/hep-th/9807080}{{\tt hep-th/9807080}}.

\bibitem{Park:1999eb}
J.~Park, R.~Rabadan, and A.~M. Uranga, {\it {N = 1 type IIA brane
  configurations, chirality and T- duality}},  {\em Nucl. Phys.} {\bf B570}
  (2000) 3--37, \href{http://xxx.lanl.gov/abs/hep-th/9907074}{{\tt
  hep-th/9907074}}.

\bibitem{Martelli:2008si}
D.~Martelli and J.~Sparks, {\it {Moduli spaces of Chern-Simons quiver gauge
  theories and AdS(4)/CFT(3)}},  {\em Phys. Rev.} {\bf D78} (2008) 126005,
  \href{http://xxx.lanl.gov/abs/0808.0912}{{\tt 0808.0912}}.

\bibitem{Hanany:2008cd}
A.~Hanany and A.~Zaffaroni, {\it {Tilings, Chern-Simons Theories and M2
  Branes}},  {\em JHEP} {\bf 10} (2008) 111,
  \href{http://xxx.lanl.gov/abs/0808.1244}{{\tt 0808.1244}}.

\bibitem{Hanany:2008fj}
A.~Hanany, D.~Vegh, and A.~Zaffaroni, {\it {Brane Tilings and M2 Branes}},
  {\em JHEP} {\bf 03} (2009) 012, \href{http://xxx.lanl.gov/abs/0809.1440}{{\tt
  0809.1440}}.

\bibitem{Jafferis:2008qz}
D.~L. Jafferis and A.~Tomasiello, {\it {A simple class of N=3 gauge/gravity
  duals}},  {\em JHEP} {\bf 10} (2008) 101,
  \href{http://xxx.lanl.gov/abs/0808.0864}{{\tt 0808.0864}}.

\bibitem{Ueda:2008hx}
K.~Ueda and M.~Yamazaki, {\it {Toric Calabi-Yau four-folds dual to
  Chern-Simons-matter theories}},  {\em JHEP} {\bf 12} (2008) 045,
  \href{http://xxx.lanl.gov/abs/0808.3768}{{\tt 0808.3768}}.

\bibitem{Hanany:2008gx}
A.~Hanany and Y.-H. He, {\it {M2-Branes and Quiver Chern-Simons: A Taxonomic
  Study}},  \href{http://xxx.lanl.gov/abs/0811.4044}{{\tt 0811.4044}}.

\bibitem{Imamura:2008qs}
Y.~Imamura and K.~Kimura, {\it {Quiver Chern-Simons theories and crystals}},
  {\em JHEP} {\bf 10} (2008) 114, \href{http://xxx.lanl.gov/abs/0808.4155}{{\tt
  0808.4155}}.

\bibitem{Lee:2007kv}
S.~Lee, S.~Lee, and J.~Park, {\it {Toric AdS(4)/CFT(3) duals and M-theory
  crystals}},  {\em JHEP} {\bf 05} (2007) 004,
  \href{http://xxx.lanl.gov/abs/hep-th/0702120}{{\tt hep-th/0702120}}.

\bibitem{Feng:2001xr}
B.~Feng, A.~Hanany, and Y.-H. He, {\it {Phase structure of D-brane gauge
  theories and toric duality}},  {\em JHEP} {\bf 08} (2001) 040,
  \href{http://xxx.lanl.gov/abs/hep-th/0104259}{{\tt hep-th/0104259}}.

\bibitem{Feng:2001bn}
B.~Feng, A.~Hanany, Y.-H. He, and A.~M. Uranga, {\it {Toric duality as Seiberg
  duality and brane diamonds}},  {\em JHEP} {\bf 12} (2001) 035,
  \href{http://xxx.lanl.gov/abs/hep-th/0109063}{{\tt hep-th/0109063}}.

\bibitem{Beasley:2001zp}
C.~E. Beasley and M.~R. Plesser, {\it {Toric duality is Seiberg duality}},
  {\em JHEP} {\bf 12} (2001) 001,
  \href{http://xxx.lanl.gov/abs/hep-th/0109053}{{\tt hep-th/0109053}}.

\bibitem{Seiberg:1994pq}
N.~Seiberg, {\it {Electric - magnetic duality in supersymmetric nonAbelian
  gauge theories}},  {\em Nucl. Phys.} {\bf B435} (1995) 129--146,
  \href{http://xxx.lanl.gov/abs/hep-th/9411149}{{\tt hep-th/9411149}}.

\bibitem{Karch:1997ux}
A.~Karch, {\it {Seiberg duality in three dimensions}},  {\em Phys. Lett.} {\bf
  B405} (1997) 79--84, \href{http://xxx.lanl.gov/abs/hep-th/9703172}{{\tt
  hep-th/9703172}}.

\bibitem{Aharony:1997gp}
O.~Aharony, {\it {IR duality in d = 3 N = 2 supersymmetric USp(2N(c)) and
  U(N(c)) gauge theories}},  {\em Phys. Lett.} {\bf B404} (1997) 71--76,
  \href{http://xxx.lanl.gov/abs/hep-th/9703215}{{\tt hep-th/9703215}}.

\bibitem{Giveon:2008zn}
A.~Giveon and D.~Kutasov, {\it {Seiberg Duality in Chern-Simons Theory}},  {\em
  Nucl. Phys.} {\bf B812} (2009) 1--11,
  \href{http://xxx.lanl.gov/abs/0808.0360}{{\tt 0808.0360}}.

\bibitem{Franco:2009sp}
S.~Franco, I.~R. Klebanov, and D.~Rodriguez-Gomez, {\it {M2-branes on Orbifolds
  of the Cone over $Q^{1,1,1}$}},
  \href{http://xxx.lanl.gov/abs/0903.3231}{{\tt 0903.3231}}.

\bibitem{Davey:2009sr}
J.~Davey, A.~Hanany, N.~Mekareeya, and G.~Torri, {\it {Phases of M2-brane
  Theories}},  {\em JHEP} {\bf 06} (2009) 025,
  \href{http://xxx.lanl.gov/abs/0903.3234}{{\tt 0903.3234}}.

\end{thebibliography}\endgroup
\bibliographystyle{JHEP}

\end{document}